\documentclass[Afour,sageh,times]{sagej}

\usepackage{moreverb,url}

\usepackage[colorlinks,bookmarksopen,bookmarksnumbered,citecolor=red,urlcolor=red]{hyperref}

\newcommand\BibTeX{{\rmfamily B\kern-.05em \textsc{i\kern-.025em b}\kern-.08em
T\kern-.1667em\lower.7ex\hbox{E}\kern-.125emX}}

\def\volumeyear{2016}

\begin{document}

\runninghead{Gontcho A Gontcho, Kikwaya Eluo and Gabor}

\title{Resurfaced 1964 VRT video interview of Georges Lema\^itre}

\author{Satya Gontcho A Gontcho\affilnum{1}, Jean-Baptiste Kikwaya Eluo\affilnum{2} and Paul Gabor\affilnum{2}}

\affiliation{\affilnum{1} Lawrence Berkeley National Laboratory, 1 Cyclotron Road, Berkeley, CA 94720, USA\\
\affilnum{2} Vatican Observatory Research Group, Steward Observatory, University of Arizona, 933 N. Cherry Avenue, Tucson, AZ 85719, USA}

\corrauth{Jean-Baptiste Kikwaya Eluo,\,\, Satya Gontcho A Gontcho}
\email{\href{mailto:jbkikwaya@arizona.eduv}{jbkikwaya@arizona.edu},\,\, \href{mailto:SatyaGontcho@lbl.gov}{SatyaGontcho@lbl.gov}}

\begin{abstract}

On December 31 2022, the Vlaamse Radio- en Televisieomroeporganisatie (VRT), the national public-service broadcaster for the Flemish Community of Belgium, recovered a video recording of a 1964 interview of Georges Lemaître. 
Up until now, that footage was thought to have been lost. 
This footage represents a unique insight into the views of the physicist often coined as the "father of the Big Bang". 
The interview was conducted in French and is available online with Flemish subtitles. 
In an effort to make this treasure broadly available, we provide in this paper some brief context, an English translation of the interview as well as the French transcript for reference. 
\end{abstract}


\maketitle

\section{Introduction}

Georges Lemaître (1894-1966) was a Professor of Physics at the Catholic University of Leuven and a Roman Catholic priest. 
He is widely known as being among the firsts to formulate a theory of the Big Bang. 
He has also independently attributed the observed recession of nearby galaxies to an expanding Universe. 
In 2018, the International Astronomical Society voted to rename Hubble Law the \textit{Hubble–Lemaître Law}. 
Georges Lemaître is undeniably one of the key physicist of the XX$^{th}$ century and an important figure of the history of astronomy \citep{Lemaitre1927, Lemaitre1931a, Lemaitre1931b}. 
A video interview of Georges Lemaître talking about his work is a historical gem. 
As such, we aim to make this recording as accessible as possible for the astronomy community, and the general public at large. 
In the next section, we give some context on the recovered footage, clarify the translation process and provide the translation of the interview.

\section{The VRT Interview}

On Saturday December 31 2022, the Vlaamse Radio- en Televisieomroeporganisatie (VRT) published an online article making available to the public a video interview of Georges Lemaître: \href{https://www.vrt.be/vrtnws/fr/2022/12/31/la-vrt-a-retrouve-dans-ses-archives-une-interview-de-1964-de-geo/}{https://www.vrt.be/fr/2022/12/31/la-vrt-a-retrouve-dans-.../}. 
This 19 minutes and 47 seconds interview was thought to have been lost after its broadcasting on Friday February 14 1964, and only a $\sim$3 minutes excerpt preserved. 
Luckily, it was recently recovered by the VRT as the video had in fact been mislabelled and therefore misclassified. 
To our knowledge, it is the only video interview of Georges Lemaître in existence. 
It took place about 2 years before his passing. 
The interview was pre-recorded at a date unknown. 
Journalist Jerome Verhaeghe conducted the interview and did the on-air introduction on the day of the broadcast. 
During the interview, Georges Lemaître mentions a conference in Namur (Belgium) that took place a couple months prior. 
On Sunday June 23 1963, Georges Lemaître held a public lecture in Namur titled \textit{"Universe \& Atom"} at the "Leuven at work: scientific research from recent years" conference organized by the Association of the Friends of Leuven. 
The proceeding from this conference was given to one of the organizers who deemed it, at the time, too technical to be published. 
In 2007, the family from the organizer passed the Namur proceeding onto philosopher of science Dominique Lambert who made it public in \cite{Lambert2007}. 
Based on available historical records, this is deemed to be the last publicly held lecture that Georges Lemaître has given. 
However, it is difficult to establish with certainty that this is the conference in Namur that Georges Lemaître refers to in the VRT interview.\\

The transcript of the VRT interview, in French, can be found in the appendix. 
The transcript was done to the best of our abilities to retain word for word what Georges Lemaître said. 
For the sake of clarity, the English translation has been very lightly edited. 
For the reader to know what to expect with regards to the fidelity of the French to English translation, let us illustrate the editing choices we have made with an example fully in English. 
If the word for word transcript read "The... hum ... to address your question properly I will.... I'll focus first of all on... you know... on aspect X", the English translation will read "To address your question properly, I will focus first of all on aspect X". 
The translation is by no means perfect. 
We tried to stick as closely as possible to what Georges Lemaître said, without adding any modern turn of phrases, but simply removing superfluous mumbling and repetitions that are common in unscripted live conversations.\\ 

In the following translation, Jerome Verhaeghe will be referred to as JV and Georges Lemaître as GLM. 
We have added \texttt{[in brackets]} some comments such as, but not limited to, pointing out part of the audio that are unintelligible, change of voice from GLM in order to add emphasis to his speech, and some clarifications on what "he"/"it"/"this" refers to when the phrases get long and the repeated use of such pronouns might confuse the reader.

\subsection{English translation} 
\,

\texttt{[VRT video interview starts.]} \\

\textbf{GLM:} It is not very easy to answer your question because there are many aspects that we can develop on. Perhaps we could begin, in order to situate the question, by recalling a text that the great French mathematician, \'Elie Cartan \texttt{[Élie Joseph Cartan, Fellow of the Royal Society]}, used at the beginning of a remarkable conference in 1927 on the theory of groups and geometry. He \texttt{[\'Elie Cartan]} alluded to the general tendency of modern physics which, he said, is repugnant to the idea of having laws \texttt{[of physics]}, that would in each region of space, involve space as a whole. And this is exactly what characterizes and opposes, in a way, a simple application of the Theory of Relativity to the theory of Fred Hoyle \texttt{[Sir Fred Hoyle, Fellow of the Royal Society]}. It is that Fred Hoyle, in presenting his Steady State theory, implies that the whole universe satisfies a group, a particular group, which satisfies space as a whole. Maybe it could be explained more clearly because I can see that it is not very clear. It could be explained more clearly by saying this: a very long time ago, before the theory of the expansion of the universe (some 40 years ago), we expected the universe to be static. We expected that nothing would change. It was an a priori idea that applied to the whole universe... \texttt{[interrupted by JV]} \\

\textbf{JV:} ...that was consistent with experiment\footnote{\texttt{[Could be understood as "empirical evidence".]}}... \texttt{[interrupted by GLM]} \\

\textbf{GLM:} No, not at all. Not at all! It was an a priori idea. For which there was no experiment. And the facts relating to the expansion of the universe made this theory inadmissible. So we realized that we had to admit change.  But those who wanted for there to be no change wanted to minimize this change. In a way, they would say: "while we can only admit that it changes, it should change as little as possible". Let it change only in scale. That everything happens on a larger scale but that it happens in the same way. And this is what was first introduced by Milne\texttt{[Edward Arthur Milne, Fellow of the Royal Society]} under the name of "cosmological principle" and later under the name of "perfect cosmological principle" and then by the very idea of the Steady State Theory. This is what you find in his letters, he says: "if you try to say there is some sort of coherent plan in the universe, I think I would agree". It is the idea that there is a plan. It's not the idea that the universe has laws and that these laws build something more or less coherent. But it is the idea that there is a plan. Obviously it's a bit of a Leibnizian \texttt{[Gottfried Wilhelm Lebiniz]} idea that the universe is made as well as possible, probably a pantheistic type of Leibniz idea. We can see that from his \texttt{[Milne]} own personal ideas, can't we? In any case, from the scientific point of view, it is the idea that there is an overall plan. Whereas the general idea of physics, as Cartan expressed it, is that the laws are local, are differential and that this set cannot be very regular. At least not as regular as Hoyle expected...\\

\textbf{JV:} ... it doesn't necessarily have to be this regular? \\

\textbf{GLM:} It doesn't have to be regular, and in any case, it should not be expected to. And that's what Hoyle senses. And he has a difficulty in admitting it. You remember, he said that when he started this theory, he thought he had to reject it. The expression he wrote down: "well nothing much happens, nothing much happens" \texttt{[...]} because, he said, there should be \textit{creation}. What does this mean... creation ? This word, creation, brings with it a whole philosophical or religious resonance that has nothing to do with the question. Behind this word, creation, what is there? There is simply that the apparition of hydrogen, as Hoyle supposes, is something quite fantastic and unexpected. That's why he used the word creation. It is absolutely unexpected. And if I had to use another imagery to express the same thing, I would say that this hydrogen appears in a totally unexpected way like a ghost \texttt{[dramatic emphasis on the word "ghost"]}. It's a kind of ghost as it would appear in castles in Scotland. To introduce a kind of ghostly hydrogen in this way would avoid the difficulty that the principle of Steady State seemed to be in opposition to the Principle of Conservation of Energy. In opposition with basically the most secure and solid thing in physics \texttt{[i.e. the Energy Conservation Principle]}. In order to maintain all this, one admits a... ghostly production of hydrogen. And what can we expect from hydrogen appearing without any physical reason, without any normal connection? One could... expect it to disappear in the same way as it appeared. So this is how this theory presents itself. As a theory imposing an assumption analogous to the apriori that you should look for a static solution ... \texttt{[unclear]} ... that you are looking for a solution with a minimum of change. Which, for my part, along with others, I am opposed to \texttt{[that static solution]} in the sense that I don't think that it is the tendency of modern physics to admit that there are global laws in the universe, absolute laws, laws that, in Hoyle's expression, would imply a "design", would imply a plan. I cannot picture things working that way.\\

\textbf{JV:} You are one of the creators of a theory that is a part, if you will, of what Hoyle calls the "Big Bang" theories, the theories of the big explosion. What is your theory, and if possible at all Monsignor\footnote{\texttt{[Honorific form of address for certain members of the male clergy (akin to the English "Your Excellency").]}}, could you explain it in just a few words?\\

\textbf{GLM:} Look, I wouldn't want to... the question is too broad for this. 
In fact, Hoyle recognizes that there are many theories that he calls "Big Bang theory", right. 
I don't know to what extent the arrows he shoots against these theories are actually hitting the mark. But as far as the oldest of these theories is concerned, the one I proposed in 1931 under the name of the \textbf{hypothesis of the primeval atom}\footnote{FR: hypothèse de l'atome primitif}, I have the impression that his arrows do not reach me at all. 
I am even a little surprised that he doesn't realize this, because we have so often had the opportunity to talk, Hoyle and me, in the most friendly way. 
If you like, I have the impression that he was mostly concerned with what he imagined the theory to be rather than what it is. 
Especially the aspect of it that I have developed. 
So I think that this theory escapes... \texttt{[unintelligible]} ... if you like, I can try... \texttt{[unintelligible]} ... I can't develop the whole theory, but I'll say how it escapes \texttt{[Hoyle's arrows]}. 
He basically says that according to his theories... I would like to find his exact text \texttt{[GLM ruffling through his sheets of paper]}: "And quite early on in the game, all the galaxies are supposed to have formed and then should therefore be.." \textit{and so on}\footnote{\texttt{[said in English by GLM.]}}. 
This is not at all how I have ever considered the theory of the primeval atom. 
There is a beginning... we may touch on other aspects at some point... there is a beginning very different from the present state of the world, a beginning in multiplicity which can be described in that it can be described in the form of the disintegration of all existing matter into an atom. 
What will be the first result of this disintegration, as far as we can follow the theory, is in fact to have a universe, an expanding space filled by a plasma, by very energetic rays going in all directions. 
Something which does not look at all like a homogeneous gas. 
Then by a process that we can vaguely imagine, unfortunately we cannot follow that in very many details, gases had to form locally; gas clouds moving with great speeds... \texttt{[interrupted by JV]}\\

\textbf{JV:} Condensations\footnote{\texttt{[Could also be translated as \textit{condensates}.]}}?\\

\textbf{GLM:} No, it is not a question of condensation. Gas. Not condensations precisely, because all the universe was... all the plasma was quasi homogeneous. 
There had to have been places where gas clouds formed which stopped the rays and in this process of stopping the rays (whatever the process by which it is done), it is obvious that the very high nuclei must have broken up, must have produced a lot of hydrogen. 
So that from the point of view of astronomical development of the whole universe, we find ourselves with distinct gaseous clouds which are almost entirely made up of hydrogen. 
Now this is the key point of Hoyle's theory: it all starts with hydrogen. 
The essential difference is whether this hydrogen is produced naturally by a reasonable physical process or, on the contrary, it is a kind of phantom hydrogen which appears with just the right amount of hydrogen to verify an a priori law. 
Obviously, we can expect that a part of these rays has escaped from this condensation process. And here we have a different element, perhaps in the theory of the primeval atom: it is that a part of the rays has escaped with almost no hydrogen and that these rays can be found... we can hope that they are found, in the cosmic rays. 
This is a theory that was put forward not only by myself, but by Regener \texttt{[Erich Rudolf Alexander Regener]} a long time ago, who called cosmic rays \textit{fossil rays} in the sense that they are the testimony of the very first ages of the world. 
And I, for my part, preferred to call them the \textit{rays of the primeval fireworks}, which are preserved in the remarkably empty space and reach us... giving us a testimony of the first ages of the world... obviously, a bit of poetry in there. 
Anyway. 
The starting point remains the same, it is hydrogen, whether it is created hydrogen or hydrogen coming from other things. Bondi \texttt{[Sir Hermann Bondi, Fellow of the Royal Society]} assures us that Hoyle's work was inspired by the Steady State theory and that it \texttt{[Hoyle' work]} might not have been produced if Hoyle had not invented the Steady State theory. 
I recognize that this is certainly a very great excuse for this \texttt{[Steady State]} theory because I have the greatest admiration for Hoyle's work. 
It brought something very first class, very solid... the way in which he and his collaborators were able to show how, from hydrogen in the stars, different \texttt{[celestial]} bodies could form; it is a first class result. 
And I would say \texttt{[a result]} almost of a much more solid level than anything we can do as a cosmogonic theory. 
So if it is really this idea of Steady State theory that made this possible, well, I would be happy about that.  
But what I have to say is that, for my part, I find it just as simple to welcome this magnificent result within the framework of a theory where hydrogen is natural hydrogen rather than to welcome it in a theory where hydrogen would be phantom hydrogen, I don't see any difference between the two.\\

\textbf{JV:} Monsignor, does the fact that the universe, according to your theory, has a beginning (at least one beginning)... does it have a religious meaning for you, a religious significance?\\

\textbf{GLM:} Well, of course it must be explained... It is quite clear that in these interviews, the use of the word creation has provoked a rather particular turn in these interviews. 
Each one of them freely, and legitimately, exposed their views... These views were presented rather from the agnostic, materialist, or rather pantheist point of view. I don't think there would be any interest... any advantage in me opposing these positions that everybody are aware of... to make a confession of my religious convictions. It wouldn't make much sense. 
The point is, to answer your question by dismissing precisely what many people expect, \texttt{[the point]} is that I am not defending the primeval atom for the sake of whatever religious ulterior motive. 
Of course, nobody knows exactly what one's psychology is, really. 
But, not only consciously I don't have this idea at all, I think that the impact of this theory in the philosophical, philosophicoreligious problem is essentially different. 
It is a point obviously a little delicate. 
I am a bit afraid to elaborate on it in a few words now. 
I elaborated on it extensively in a conference that I gave a few months ago in Namur and that was on your airwaves, but on the French airwaves\footnote{Belgian Radio-television of the French Community (R.T.B.F.)} and it is precisely that if my theory is correct, it makes the philosophical problem of creation disappear, in a way. 
This is, moreover, what was developed in the preface of my little book on the primeval atom by Professor Gonseth \texttt{[Ferdinand Gonseth]} from Zurich who showed that there was a certain philosophical resonance of my theory in the sense that it made disappear some antinomies posed by Kant \texttt{[Immanuel Kant]}. \\

\textbf{JV:} If I'm not mistaken, Monsignor, did you not warn the religious authorities even against... \texttt{[interrupted by GLM]}\\

\textbf{GLM:} Look, first of all I am far too respectful of the religious authorities who know well enough what they have to do. 
But certainly I would warn... I have always warned my colleagues, well-meaning spirits, who would like to take argument on this. 
But let's go a little bit more into the substance of the question. 
When one poses the problem of the beginning of the world, one is almost always faced with a rather essential difficulty: to ask oneself, why did it begin at that moment?  
Why didn't it start a little earlier? 
And in a certain sense, why wouldn't it have started a little earlier? So it seems that any theory that involves a beginning must be unnatural. 
To say "we decide at this point that it begins"... This is what was expressed by saying: "it is made of nothing". 
That is to say that we expected it to come from something; and we say "it doesn't come from this something, it's made of nothing". 
Well... the point of view I'm coming to is quite different. 
That is, the beginning is so unimaginable, so different from the present state of the world that such a question does not arise. 
And even more than that. 
This beginning is the beginning of multiplicity. 
The fundamental idea is... I can't develop it with more details now... it is the beginning of multiplicity. 
It is the idea that the universe, which exists in quanta, in packets of determined energy, begins with a single quantum, or a very small number of quanta, so that it is impossible to wonder from what it would come, from what it would have been divided from. 
The whole development of entropy is that quanta divide themselves, develop, etc. At the beginning, if there is only one, we cannot ask ourselves where it comes from. 
Then the question does not arise to say that it comes from nothing. It is a background of space-time for which no problem arises. 
Or if you want, when one holds oneself as the spiritualist, with the idea that it comes from God, etc. ... well, one would like to take God in default for this 'initial flick' as Laplace \texttt{[Pierre-Simon, Marquis de Laplace]} said. 
Well it doesn't hold, because the beginning... the bottom of the space-time is so different from all our conceptions that there is no more problem. 
And then obviously for an atheist, everything/anything cannot remain, I cannot continue to speak if God doesn't support me in the existence, that's for sure, isn't it? 
But that's nothing, that's the general stance of christian philosophy. 
But there's nothing special about the beginning. And the beginning is not a place where you would touch God as a hypothesis, where if you like, I'll talk about Laplace's initial flick, since we're now talking about conferences in English... I recall Jeans \texttt{[Sir James Hopwood Jeans, Fellow of the Royal Society]} words "the finger of God agitating the ether" \texttt{[dramatic voice]}, that was the beginning. 
Well, that's not... that's not a pleasant idea for a religious mind. 
It's an idea that brings God down into the realm of primary causes, and I think one of the contributions that a theory like mine can make is to avoid just such difficulties.\\

\textbf{JV:} So, to state things plainly, you refuse to accept the idea that God should explain the movement of galaxies. \\

\textbf{GLM:} Of course, it goes without saying! Absolutely. That is to say if God supports the galaxies, he acts as God, he does not act as a force that would contradict everything. It's not Voltaire's watchmaker who has to wind his clock from time to time, isn't it... [laughs]. There!\\

\textbf{JV:} Thank you very much, Monsignor. \\

\texttt{[VRT video interview ends.]} \\

\section{Remarks on the term \textit{"initial flick"}}
\,

It seems that the term \textit{chiquenaude}, translated as \textit{flick}, is being misattributed to Laplace in this interview. 
The term was used by Blaise Pascal to refute René Descartes’ form of theism. 
In his book \textit{Thoughts} \citep{Pascal}, Pascal says: "Je ne puis pardonner à Descartes: il aurait bien voulu, dans toute sa philosophie, pouvoir se passer de Dieu ; mais il n’a pu s’empêcher de lui faire donner une chiquenaude pour mettre le monde en mouvement."  
In English: "I cannot forgive Descartes: he would have liked, in all his philosophy, to be able to do without God; but he could not help to have him give a flick to set the world in motion."\\


\begin{acks}

The authors thank the Vlaamse Radio- en Televisieomroeporganisatie (VRT), the national public-service broadcaster for the Flemish Community of Belgium, for access to the 1964 recovered video interview of Georges Lemaître. In particular, Kathleen Bertrem (VRT Archive \& Communication departments) for her availability. Our thanks also go to Dominique Lambert for his prompt return on our draft and sharing his experience and expertise on Georges Lemaître with us.
SGG is supported by the Director, Office of Science, Office of High Energy Physics of the U.S. Department of Energy under Contract No. DE-AC02-05CH1123. JBKE and PG are supported by the Vatican and the Vatican Observatory Research Group.

\end{acks}

\section{Appendix: French transcript}\label{sec:appendix}
\,

\texttt{[Vidéo de la VRT commence.]} \\

\textbf{GLM:} Et bien ce n'est pas très facile de répondre à cette question. Parce qu'il y a beaucoup d'aspects que l'on peut développer. Mais peut-être pourrions-nous commencer pour situer la question par rappeler un texte que le grand mathématicien français, \'Elie Cartan, avait employé en débutant une conférence remarquable, il fut en 1927, sur la théorie des groupes et la géométrie. Et voici ce qu'il dit. Il fait allusion à la tendance générale de la physique moderne qui répugne, dit-il, à l'idée de soumettre des lois a priori, faisant intervenir dans chaque région de l'espace, l'espace tout entier. Et c'est exactement cela qui caractérise et qui oppose, en quelque sorte, une simple application de la théorie de la relativité à la théorie de Fred Hoyle. C'est que, Fred Hoyle, en présentant sa Steady State Theory\footnote{FR: théorie de l'état stationnaire} implique que l'univers tout entier satisfait à un groupe, un groupe particulier, qui vérifie tout l'ensemble de l'espace. Mais peut-être pourrait-on encore l'expliquer plus clairement parce que je vois que ce n'est pas très clair ainsi. On pourrait peut-être l'expliquer plus clairement en se disant ceci : avant, il y a très longtemps, avant la théorie de l'expansion de l'univers il y a une quarantaine d'années, et bien on s'attendait à ce que l'univers soit statique, à ce que que rien ne change. Et bien, c'était au fond une idée a priori qui s'appliquait à tout l'ensemble de l'univers… \texttt{[interrompu par JV]} \\

\textbf{JV:} … qui  correspondait à l'expérience \texttt{[interrompu par GLM]}\\

\textbf{GLM:} Non, pas du tout. Pas du tout! C'était une idée a priori. Pour laquelle il n'y avait aucune expérience. Et les faits relatifs à l'expansion de l'univers ont rendu cette théorie inadmissible. Alors on s'est rendu compte qu'il fallait admettre changement. Mais ceux qui tenaient à ce qu'il n'y ait pas de changement ont voulu minimiser ce changement. Et on dit "Et bien, il faut bien admettre que ça change, mais que ça change le moins possible. Que ça change simplement d'échelle. Que tout se reproduise mais à une échelle un peu plus grande. Que tout se reproduise de la même façon." Et c'est cela qui a d'abord été introduit par Milne sous le nom de \textit{principe cosmologique} et plus tard \textit{principe cosmologique parfait}. Et ensuite l'idée même de la Steady State Theory. C'est bien cela, n'est-ce-pas, que nous pouvons trouver dans ses textes: il dit: “if you try to say there is some sort of coherent plan in the universe, I think I would agree”. C'est l'idée qu'il y a un plan. Ce n'est pas l'idée que l'univers a des lois et que ces lois bâtissent plus ou moins quelque chose de plus ou moins cohérent. \texttt{[défaut bande audio]} C'est l'idée qu'il y a un plan. \'Evidemment c'est dans le fond une idée du type de Leibniz; que l'univers est fait pour le mieux possible, d'un type de Leibniz probablement... probablement panthéiste. Nous le voyons bien d'après les idées personnelles qu'il développe, n'est-ce-pas? Mais en tout les cas, du point scientifique, une idée qu'il y a un plan d'ensemble. Tandis que l'idée générale de la physique comme l'exprimait Cartan, c'est que les lois sont locales, sont différentielles et que l'ensemble ne peut pas avoir une très grande régularité comme l'attend Hoyle. \\

\textbf{JV:} Ne doit pas nécessairement avoir cette régularité ? \\

\textbf{GLM:} Ne doit pas nécessairement l'avoir et qu'on ne doit pas l'attendre. Et d'ailleurs c'est bien ce que Hoyle sent, c'est qu'il a une difficulté à l'admettre. Vous vous rappelez qu'il me disait que lorsqu'il a commencé cette théorie, il a cru devoir la rejeter. Et l'expression qu'il notait : "well nothing much happened. nothing much happened", n'est-ce pas, pour un temps. Parce que, dit-il, il devrait y avoir création. Mais qu'est-ce que ça veut dire, ce mot création ? Ce mot création, au fond, ça entraîne toute une résonnance philosophique ou religieuse qui n'a rien à avoir avec la question. Derrière ce mot de création, qu'est-ce qu'il y a ? Il y a tout simplement que l'apparition de l'hydrogène, comme la suppose Hoyle, est quelque chose de tout à fait fantastique et inattendu. C'est pour ça qu'il a employé le mot création. C'est absolument inattendu. Et si je devais employer une autre image pour exprimer la même chose, je dirais que cet hydrogène apparaît de manière totalement inattendue comme un fantôme \texttt{[emphase dramatique sur ce dernier mot]}. C'est une sorte de fantôme tel qu'ils apparaissent, paraît-t-il, dans les châteaux d'\'Ecosse. Et introduire une sorte d'hydrogène fantomal ainsi pour... pour éviter la difficulté qui se présentait, la difficulté qui se présentait, c'est que le principe de Steady State paraissait en opposition avec la conservation d'énergie. Avec au fond ce qu'il y a de plus sûr, de plus solide en physique. N'est-ce-pas ? Et bien pour maintenir tout de même tout cela, on admet une production d'hydrogène absolument fantomale. Et qu'est-ce que l'on peut attendre d'un hydrogène apparaissant ainsi; sans aucune raison physique, sans aucune connexion normale ? On pourrait tout au plus attendre qu'il disparaisse comme il est apparu. Alors voilà donc comment cette théorie se présente, au fond. Comme une théorie imposant un \textit{a priori}. Un apriori analogue à l'apriori que vous pouvez avoir de chercher une solution statique, de chercher une solution avec le minimum de changement. Et contre lequel, pour ma part, et avec d'autres, je m'oppose dans ce sens que je ne pense pas que \texttt{[... cherche ses mots ...]} ça soit la tendance de la physique moderne d'admettre qu'il y a dans l'univers des lois globales, des lois absolues, des lois qui, dans l'expression de Hoyle, impliqueraient un design, impliqueraient un but, un plan, n'est-ce-pas? Ce n'est pas comme ça enfin que je puis me représenter les choses.\\

\textbf{JV:} Vous êtes un des créateurs d'une théorie qui est une partie, si vous voulez, dans l'ensemble de ce que Hoyle appelle les "Big Bang Theory". Donc les théories de la grande explosion. Quelle est votre théorie, si c'est possible bien entendu Monseigneur de l'expliquer en quelques mots seulement ?	\\

\textbf{GLM:} Mais écoutez, je ne voudrais pas... mmh, la question est trop vaste comme ceci. Mais enfin, Hoyle reconnaît qu'il y a beaucoup de théories qu'il appelle "Big Bang theory", n'est-ce pas... \texttt{[inaudible]} Moi je ne sais pas jusqu'à quel point les flèches qu'il décoche contre ces théories, réellement, atteignent leur but. Mais pour ce qui concerne la théorie la plus ancienne parmis ces théories, celle donc que j'ai proposé en 1931 sous le nom d'\textbf{hypothèse de l'atome primitif}, bah j'ai l'impression que ses flèches ne m'atteignent pas du tout. Et je suis même un peu étonné qu'il ne s'en rende compte pas tellement parce qu'enfin nous avons tellement souvent eu l'occasion de causer, Monsieur Hoyle et moi, de la manière la plus amicale, n'est-ce-pas... \texttt{[inaudible]} ... j'ai l'impression qu'il s'est surtout attaché à ce qu'il s'imaginait de ces théories plutôt que, et surtout l'aspect particulier d'ailleurs que j'ai toujours développé. Je pense donc que cette théorie échappe pour l'instant ... je ne vais pas développer l'ensemble de la théorie, mais je voudrais dire, essentiellement, en quoi elle échappe. Il dit essentiellement que d'après ses théories... j'aimerais retrouver son texte exact \,\texttt{[fouille dans ses feuilles de papier]}\, "And quite early on the game, all the galaxies are supposed to have formed and then should therefore be.. " \textit{and so on}\footnote{\texttt{[inaudible]}}. Mais ce n'est pas du tout ainsi que j'ai jamais considéré la théorie de l'atome primitif. Il y a un commencement, nous y reviendrons peut-être à d'autres aspects, un commencement très différent de l'état actuel du monde, un commencement en multiplicité qui peut être décrit pour autant qu'on peut le faire sous forme de la désintégration de toute la matière existante sous forme d'un atome. Quel va être le premier résultat de cette désintégration pour autant qu'on peut suivre la théorie, et bien c'est d'avoir un univers, un espace en expansion rempli par un plasma, par des rayons très énergiques allant dans tous les sens. Quelque chose qui ne ressemble pas du tout à un gaz homogène. Alors par un procédé que l'on peut vaguement imaginer, mais malheureusement qu'on ne peut pas suivre fort en détails, il a dû se former localement, il a dû se former des gaz par places, de nuées gazeuses animées de grandes vitesses... \texttt{[interrompu par JV]}\\

\textbf{JV:} Des condensations ?\\

\textbf{GLM:} Pas... hm il ne s'agit pas de condensations. Des gaz! Pas précisément des condensations, parce que tout l'univers était, tout le plasma était ainsi de manière quasi homogène. Il a dû se former, par places, du gaz, qui a alors arrêté les rayons et dans ce processus d'arrêt (quel que soit le détail dans lesquel ça se fait), il est évident que les noyaux très élevés ont dû se briser, finalement, ont dus produire beaucoup d'hydrogène. De sorte que, au point de vue du développement vraiment astronomique de tout l'univers, on se trouve devant des nuées gazeuses distinctes qui sont presque entièrement de l'hydrogène. Or c'est là l'essentiel de la théorie de Hoyle, c'est de commencer avec de l'hydrogène. L'essentiel, n'est-ce pas, la différence, c'est de savoir si cet hydrogène s'est produit naturellement par un processus physique raisonnable, ou au contraire c'est une sorte d'hydrogène fantomal qui arrive tout juste avec la quantité voulue pour vérifier une loi posée a priori. \'Evidemment on peut s'attendre à ce qu'une \texttt{[inaudible]} partie de ces rayons ait échappé à ce processus de condensation. Et là, nous avons un élément différent peut-être dans la théorie de l'atome primitif, c'est qu'une partie des rayons ont échappé, et ne comptant presque pas d'hydrogène, et que ces rayons peuvent se retrouver.. et que l'on peut espérer qu'ils se trouvent, dans les rayons cosmiques. C'est une théorie qui a été avancée non seulement par moi-même, mais par Regener il y a bien longtemps qui les appelaient... il appelait les rayons cosmiques des rayons fossiles; dans ce sens qu'ils sont le témoignage des tous premiers âges du monde et que, pour ma part, je préférais appeler les rayons du feu d'artifice primitif, qui se sont conservés dans l'espace remarquablement vide et nous parviennent en nous témoignant, en nous donnant un témoignage des premiers âges du monde... évidemment un peu de poésie là-dedans. Quoiqu'il en soit, n'est-ce pas... le point de départ reste le même c'est l'hydrogène, que ce soit de l'hydrogène créé, que ce soit de l'hydrogène provenant d'autre chose. Bondi nous assure que les travaux de Hoyle ont été inspirés par la "Steady State theory" et qu'ils [les travaux de Hoyles] n'auraient peut-être pas été produits si Hoyles n'avait pas inventé la Steady State Theory. Je reconnais que c'est certainement au moins une très grande excuse pour cette théorie. Parce que j'ai la plus grande admiration pour ces travaux de Hoyle qui ont apporté quelque chose de tout premier plan, de tout à fait solide, que la manière dont lui-même, avec des collaborateurs, a pu montrer comment à partir de l'hydrogène pouvait se former dans les étoiles les différents corps est un résultat de tout premier plan et je dirais presque d'un niveau beaucoup plus solide que tout ce que nous pouvons faire comme théorie cosmogonique. Donc si réellement c'est cette idée de Steady State Theory qui a rendu cela possible, ben, je m'en réjouirais.  Mais ce que je dois dire, c'est que, pour ma part, je trouve tout aussi simple d'accueillir ces magnifiques résultats dans le cadre d'une théorie où l'hydrogène est de l'hydrogène naturel plutôt que de l'accueillir dans une théorie où de l'hydrogène serait de l'hydrogène fantomal, je ne vois pas de différence entre les deux.\\

\textbf{JV:} Monseigneur, est-ce que le fait que l'univers, selon votre théorie, a un commencement (au moins un commencement), est-ce que cela a pour vous un sens religieux, une signification religieuse ?\\

\textbf{GLM:} Et bien, il faut naturellement vous expliquer, il y a, il est bien clair que dans ces interviews, l'emploi du mot création a provoqué un tour assez particulier à ces interviews.  Chacun d'entre eux a très librement et très légitimement exposé ses vues.. ces vues se présentaient plutôt sous une forme agnostique, matérialiste, ou plutôt panthéiste. Je ne pense pas qu'il y aurait le moindre intérêt, le moindre avantage que j'oppose à ces positions que tout le monde connaît, que je fasse ici une profession de mes convictions religieuses, ca n'aurait pas beaucoup de sens. Aussi, je pense que le point de vue est plutôt de répondre à votre question en écartant justement ce que beaucoup de gens imaginent: c'est que je défends l'atome primitif, je ne sais pas moi avec quelle idée, quelle arrière-pensée religieuse, n'est-ce-pas. Naturellement, personne ne sait exactement quelle est sa psychologie et ce qu'il fait réellement. Mais, non seulement consciemment je n'ai pas du tout cette idée, n'est-ce-pas... Mais même, n'est-ce-pas, je pense que l'impact de cette théorie dans le problème philosophique... philosophico-religieux, est essentiellement différent. C'est un point évidemment un peu délicat, j'ai un peu peur de développer en quelques mots maintenant. Je l'ai développé très largement dans une conférence que j'ai faite il y a quelques mois à Namur et qui a été sur vos ondes, mais à l'antenne fran\c{c}aise\footnote{Radio-télévision belge de la Communauté française (R.T.B.F.)}, n'est-ce-pas, et c'est justement que si ma théorie est correcte, elle fait en quelque sorte disparaître le problème philosophique de la création. C'est d'ailleurs ce qui a été développé dans la préface de mon petit livre sur l'atome primitif par le Professeur Gonseth de Zurich, n'est-ce pas, qui a justement, avec raison, montré qu'il y avait une certaine résonance philosophique de ma théorie en ce sens qu'elle faisait disparaître certaines des antinomies posées par Kant. \\

\textbf{JV:} Si je ne me trompe pas, Monseigneur, est-ce que vous n'avez pas mis les autorités religieuses même en garde contre \texttt{[interrompu par GLM]} \\

\textbf{GLM:} Non écoutez, d'abord je suis beaucoup trop respectueux des autorités religieuses qui savent bien ce qu'elles ont à faire, certainement pas les autorités religieuses, mais certainement je mettrais, j'ai toujours mis en garde mes collègues, des esprits bien intentionnés, n'est-ce-pas, qui voudraient prendre argument de cela. Mais je crois qu'il vaut tout de même mieux aller un peu plus dans le fond de la question: pourquoi s'en est-il pas \texttt{[ininteligible]} ? Et bien, lorsqu'on se pose le problème du début du monde, on se trouve presque toujours devant une difficulté assez essentielle: de se demander pourquoi ça commence-t-il à ce moment-là ?   Pourquoi ça n'aurait-il pas commencé un peu plus tôt ? Pourquoi.... Et en un certain sens, pourquoi ça n'aurait-il commence un peu plus tôt? De telle sorte qu'il semble que toute théorie qui implique un commencement doit être quelque chose de peu naturel. De dire “on décide à ce moment là que ça commence”. C'est ce qui s'exprimait en disant : "c'est fait de rien". C'est-à-dire qu'on s'attendait à ce que ça provienne de quelque chose; et on dit “ça ne provient pas de ce quelque chose, c'est fait de rien". Eh bien… le point de vue auquel j'arrive est tout différent. C'est-à-dire que le commencement est tellement inimaginable, tellement différent de l'état actuel du monde qu'une pareille question ne se pose pas. Et même, plus que cela. Ce commencement est le commencement de la multiplicité. L'idée fondamentale c'est -- et cela, et je ne peux pas le développer avec quelque peu de détails maintenant. Mais enfin l'idée, c'est le commencement de la multiplicité, c'est que l'univers qui existe en quanta, en paquets d'énergie déterminée, commence avec un seul quantum, ou un tout petit nombre de quanta de telle sorte qu'il est impossible de se demander de quoi il proviendrait, de quoi il serait divisé. Tout le développement de l'entropie, c'est que les quanta se divisent, se développent, etc. Au départ s'il n'y en a qu'un, on ne peut pas se demander d'où il vient. Et alors la question ne se pose pas de dire qu'il vient de rien. C'est un fond de l'espace-temps pour lequel aucun problème ne se présente. Ou si vous voulez, lorsqu'on se tient comme le spiritualiste a l'idée que tout provient de Dieu, etc. Et bien, on voudrait prendre en quelque sorte Dieu en défaut pour dire "Et bien là on le touche dans cette chiquenaude initiale" comme disait Laplace, ou des choses ainsi... Et bien il échappe, il échappe parce que le début, le fond de l'espace-temps est tellement différent de toutes nos conceptions qu'il n'y a plus de problème. Et alors évidemment pour un athéiste, n'est-ce pas, tout/rien ne peut subsister, je ne pourrais pas continuer à parler si Dieu ne me soutenait dans l'existence, c'est bien sûr, n'est-ce pas ? Mais il n'y a rien, ça, c'est l'attitude générale de la philosophie chrétienne. Mais il n'y a rien de spécial pour le commencement. Et le commencement n'est pas un endroit où l'on toucherait Dieu comme une hypothèse, où si vous voulez, je vais parler de la chiquenaude initiale de Laplace, puisque nous parlons maintenant des conférences faites en anglais... Je rappelle les mots de Jeans "the finger of God agitating the ether" \texttt{[voix dramatique]}, voilà le commencement. Et bien, cela, n'est pas, ce n'est pas une idée agréable pour un esprit religieux. C'est une idée qui fait descendre Dieu dans le domaine des causes premières et je crois qu'une des contributions que peut apporter une théorie telle que la mienne, c'est d'éviter justement ces difficultés. \\

\textbf{JV:} Donc, vous vous refusez à l'idée que Dieu devrait expliquer le mouvement des galaxies pour le dire très concrètement. \\

\textbf{GLM:} \'Evidemment, c'est clair ça! C'est clair, c'est-à-dire que si, Dieu soutient les galaxies, mais il agit en Dieu. Il n'agit pas comme une force qui viendrait en contredire tout. Ce n'est pas l'horloger de Voltaire qui doit de temps en temps remonter son horloge, n'est-ce pas… \texttt{[Rires]}. Voilà.\\

\textbf{JV:} Merci beaucoup Monseigneur.\\ 

\texttt{[Vidéo de la VRT se termine.]} 
\end{document}